# Using Facebook for Image Steganography

Jason Hiney[1], Tejas Dakve[1], Krzysztof Szczypiorski[2], Kris Gaj[1]

[1]George Mason University
Fairfax, VA, United States of America
jhiney@gmu.edu, tejasdash02@gmail.com, kgaj@gmu.edu
[2]Warsaw University of Technology
Warsaw, Poland
ksz@tele.pw.edu.pl

*Abstract*—Because Facebook is available on hundreds of millions of desktop and mobile computing platforms around the world and because it is available on many different kinds of platforms (from desktops and laptops running Windows, Unix, or OS X to hand held devices running iOS, Android, or Windows Phone), it would seem to be the perfect place to conduct steganography. On Facebook, information hidden in image files will be further obscured within the millions of pictures and other images posted and transmitted daily. Facebook is known to alter and compress uploaded images so they use minimum space and bandwidth when displayed on Facebook pages. The compression process generally disrupts attempts to use Facebook for image steganography. This paper explores a method to minimize the disruption so JPEG images can be used as steganography carriers on Facebook.

## I. INTRODUCTION

Steganography comes from "covered writing" in Greek and means hiding one piece of information in another piece of information (the carrier). Reference [1] gives writing directly on the wooden backing of wax tablets before the beeswax had been applied, hiding information on the human body, using microdots, and using invisible ink as examples of early steganography. The computer age brings many more opportunities for steganography. Examples of digital steganography include hiding information in digital images, replacing the least significant bits of voice-over-IP transmissions, and modifying network packet structures or timing relations (network steganography).

This paper focuses on hiding information in Joint Photographic Experts Group (JPEG) [2] images then uploaded and posted to Facebook as a means of distribution. Input file formats to Facebook are manipulated in an attempt to manage the compression problem. Section II of the paper summarizes previous related work. Section III details our efforts to preprocess JPEG carrier images so that hopefully the Facebook compression algorithm performs very little compression that will disrupt steganography. Section IV describes our efforts to pick the best tool to conduct success rate testing in section V. Section VI deals with steganalysis, first in general and then applied to the project. Section VII provides concluding remarks and discusses possible future work.

## II. RELATED WORK

### A. A Forensic Analysis of Images on Online Social Networks

Castiglione, Cattaneo, and De Santis [3] analyzed how several popular online social networks process uploaded images and what changes are made to the published images. Of particular interest to our application is which image formats Facebook accepts for upload and how it transforms each file type.

### B. Stegobot: A Covert Social Network

Nagaraja, Houmansadr, Piyawongwisal, Singh, Agarwal, and Borisov [4] proposed Stegobot, a bot network utilizing Facebook image steganography to provide covert communication channels. A database of 116 different images was used to determine the maximum JPEG resolution not altered by Facebook image processing. Then images were resized below the maximum Facebook constraint, the YASS steganography scheme was utilized to embed the bot communications, the carrier images were uploaded to Facebook, and the YASS detector was used to extract the communications. In our experimentation, we also use Facebook to conduct image steganography and manipulate JPEG carrier images to minimize the changes made in Facebook image processing. However, we work with higher resolution JPEG carrier images and conduct experimentation on multiple steganography programs. In addition, our goal is to conceal and transmit whole text and image files rather than to conceal and transmit bot network communications, and our results therefore consist of success rates for transmitting various size text and image files using steganography on Facebook.

### C. Secretbook

Campbell-Moore developed the Secretbook plug-in to hide text messages up to 140 characters in JPEG images on Facebook using the Google Chrome browser. Beckhusen's article [5] does an excellent job framing the Facebook steganography problem and explaining Campbell-Moore's solution. When one uploads an image to Facebook, the image is automatically compressed. If there is steganography in the image, Facebook garbles it. The Secretbook algorithm automatically compresses a JPEG image as Facebook would and then adds the hidden steganography data. The algorithm also adds redundancy so any remaining distortion can be corrected by reconstruction from the copies. Our approach is very similar to the first step of Campbell-Moore's algorithm. We determine which image format Facebook alters the least and preprocess images in that format to minimize the chances that Facebook will corrupt the loaded carrier file to the extent that steganography is "uncovered" or secret message recovery fails. However, we will also show that it is possible to use JPEG carrier images to transmit hidden text messages longer than 140 characters and hidden images using Facebook.

III. PRELIMINARY EXPERIMENTS

*A. Selection of Carrier File Type*

In this first phase of experimentation, we uploaded almost 100 JPEG, bitmap (BMP) [6], graphics interchange format (GIF) [7], portable network graphics (PNG) [8], and tagged image file format (TIFF) [9] files to Facebook and then downloaded them, observing changes to file types and sizes. Table I summarizes characteristics of these file types.

Most of the uploads were to the Facebook main page, but we also did some uploads to the profile picture. We observed that Facebook accepts JPEG, PNG, GIF, and TIFF files but does not accept BMP files. We observed that Facebook converts uploaded PNG, GIF, and TIFF images to JPEG File Interchange Format (JFIF) images [10]. (JFIF is a standardized format for exchanging JPEG files consistent with the JPEG Interchange Format [JIF].) We observed that Facebook converts uploaded JPEG/Exchangeable Image file Format (EXIF) images [11] to JPEG/JFIF images. (EXIF is a standard format used by cameras and similar devices that contains additional metadata such as camera model, date and time, camera settings, etc., and a thumbnail for viewing on a camera screen.) We observed that Facebook compresses uploaded JPEG/JFIF images.

Kessler [12] explains how JPEG uses discrete cosine transforms in a lossy compression scheme quite different from the pixel-by-pixel lossless compression scheme used in other image file formats. Given that Facebook converts all uploaded image types to JPEGs and that JPEG compression is very different from compression in other image types, we reasoned that JPEGs offer the best chance of uploading an image type to Facebook that will be unchanged or minimally changed by compression. We observed upload to download file size ratios ranging from (0.83 to 156) for JPEGs.

*B. Finding and Applying Standard Image Download Resolutions*

In this second phase of experimentation, we first uploaded JPEGs of different resolutions to Facebook and then downloaded them, observing changes to resolution. We uploaded to a Facebook album with the high quality setting to maximize the corresponding downloaded file's size and capacity to hold hidden information. We observed that JPEG images downloaded from Facebook are generally 2048 * yyyy resolution or 960 * yyy resolution when the upload resolutions are >=960 * yyy (yyyy and yyy are the numbers of pixels on the shorter sides of the images). We observed that the upload JPEG resolutions were generally not changed upon download when the numbers of pixels on the longer sides of the images were <= 960. Considering that high image resolutions were needed to maximize the steganography payload, we concluded that JPEG images with 2048 * yyyy (hereafter 2048 resolution) and 960 * yyy (hereafter 960 resolution) resolutions were the most interesting candidates as carriers for Facebook steganography.

TABLE I. OVERVIEW OF IMAGE FILE TYPES

| BMP | • Very old Microsoft uncompressed proprietary format |
|---|---|
| GIF | • Limited to 256 colors<br>• Used for fast-loading web graphics<br>• Not suitable for images with continuous color like photographs, but well suited for simple images with solid areas of color |
| JPEG | • Most popular image file type<br>• Used to store & display images on web sites and in cameras<br>• Supports a full spectrum of colors<br>• Compatible with the vast majority of devices & programs<br>• Can be compressed to save storage space and transmission time (lossy compression) |
| PNG | • Small files that maintain original image quality<br>• Supports a full spectrum of colors and transparency<br>• Suitable for graphics image files like logos and infographics<br>• Not compatible with all software |
| TIFF | • Suitable for a bitmap image that may be edited<br>• No compression as intended to preserve quality<br>• Produces big files and thus not suitable for web graphics |

In the second part of this phase, we began manipulating JPEG files to see if we could produce download resolutions and file sizes similar to the upload resolutions and file sizes. The theory was that in this situation, Facebook might be doing very little, if anything, to transform the uploaded JPEGs and that they might then serve as good steganography carriers. So in phase 2 part 2 we took JPEG/EXIF and JPEG/JFIF sample images and used Nikon View NX2 to convert them to 2048 and 960 resolutions. We also experimented with different View NX2 compression ratios to see which ones would produce upload to download file size ratios closest to 1.0 when uploading to a Facebook album with high quality setting and then downloading from Facebook. We found that JPEG test images resized to 2048 resolution and then compressed using the "good compression" setting or the "highest compression" setting in View NX2 produced upload to download ratios closest to 1.0 (in the 1.03 to 1.45 range). This range is much closer to 1.0 than the 24 to 156 compression ratios observed in phase 1 for JPEGs with >=2048 resolution. We found that JPEG test images resized to 960 resolution and then compressed using the "good compression" setting or the "highest compression" setting in View NX2 produced upload to download ratios in the 1.68 to 2.57 range. This range is much closer to 1.0 than the 4.67 to 37.4 compression ratios observed in phase 1 for JPEGs with resolutions <2048 but >=960.

We also observed that in some cases JPEGs uploaded to Facebook in the 2048 resolution downloaded in the 960 resolution. This happened when downloading to laptops using Internet Explorer but not when downloading to laptops using Google Chrome. It did not occur when downloading to a desktop. We concluded that an upload at 2048 resolution would likely fail to successfully transmit steganography with a subsequent download at 960 resolution because the carrier file would be drastically compressed. However, both the 960 and 2048 resolutions are still valid for conducting Facebook steganography. The larger resolution offers a compression ratio closer to 1.0 and should offer more hidden payload, but the smaller resolution offers a consistent download resolution across differing computer platforms and download browsers. Because each situation offers a good characteristic for conducting steganography, and because steganography is possible under both conditions, we decided to continue testing both 2048 and 960 resolution images.

### C. Achieving 1.0 Upload to Download File Size Ratio

In the third phase of experimentation, our goal was to further develop our set of JPEG test images so that when they were uploaded to a Facebook album using the high quality setting, the upload file size to download file size ratio would be at 1.0 or very close to 1.0. At this point our test images had already been resized, converted, and compressed by NX2; uploaded to a Facebook album; and downloaded from Facebook. We reasoned that uploading them to Facebook again and downloading them from Facebook again might bring them to the 1.0 target ratio. We chose the ten 2048 resolution JPEG images with the lowest phase 2 upload to download ratios and the ten 960 resolution JPEG images with the lowest phase 2 upload to download ratios and uploaded them to a Facebook album with the high quality setting again. The 2048 resolution downloads were conducted using the desktop, so resizing to 960 resolution would not occur. The 2048 resolution downloads yielded ratios in the 0.99 to 1.00 range. The 960 resolution downloads yielded ratios in the 1.00 to 1.06 range. With the ratios at or very close to 1.0 for these carrier test files, we were ready to begin steganography testing.

### IV. STEGANOGRAPHY PROGRAM TESTING

### A. Procedure

In this fourth phase of experimentation, we created short text files and used various steganography programs to hide them in the twenty test carrier JPEG files produced by phase 3. We then uploaded the carriers to Facebook using the high quality setting, downloaded them from Facebook, and attempted to recover the hidden information. Table II shows the programs tested and provides an overview of each program's salient features.

TABLE II. STEGANOGRAPHY PROGRAMS TESTED AND SALIENT FEATURES

| Program | Operating System(s) | Interface | Additional Features |
|---|---|---|---|
| Open Puff | Windows | GUI | multiple passwords, hiding across multiple carriers, encryption, decoy |
| Outguess Rebirth | Windows | GUI | encryption |
| F5 | Java capable | command line | |
| JP Hide & Seek | Windows, Linux | GUI for Windows | |
| Steghide | Windows, Linux | command line | |
| Steg | Windows, Linux, OS X | GUI | encryption |
| Our Secret | Windows | GUI | |
| Incognito | Android | GUI | |
| Stegan-ography | Android | GUI | |

### B. Results

In each experiment, we tested the ability of the program to recover the hidden text prior to upload to Facebook and then proceeded with Facebook testing.

Open Puff did not work with our test carriers. Any JPEG images previously compressed by Facebook cause an "unsupported carrier format" error. We were surprised to see Open Puff, the best steganography software according to many users, failing to work with images already compressed by Facebook. We tested and found that it does work with JPEG images not processed by Facebook.

Outguess Rebirth did not work with our test carriers. The program has an insert/extract button that changes from "insert file" to "extract file" when it detects that a carrier is hiding information. Facebook compression alters the carrier in such a way that it does not detect data hidden before upload to Facebook. In all 20 test cases the program offered the "insert file" button and not the "extract file" button when the carrier downloaded from Facebook was loaded into the program.

F5 did not work with our test carriers. In all 20 test cases the program halted when the extraction was requested**.**

We achieved a 50% success rate with JP Hide & Seek. Ten of the 20 experiments were successful. We received "passphrase wrong" errors in 8 of the 10 failed experiments. In one case, the text was recovered, but some characters were changed.

We achieved a 15% success rate with Steghide. Three of the 20 experiments were successful. We received "could not extract with that passphrase" errors in 13 of the

17 failed experiments. We also had 2 cases where the text was recovered, but some characters were changed.

We achieved a 20% success rate with Steg. Four of the twenty experiments were successful. The program crashed in 11 of the failed experiments. There were also 5 "extraction failed" errors.

Our Secret did not work with our test carriers. In all 20 experiments we received a "The file hides no data!" error.

Incognito did not work with any JPEG files (even prior to Facebook compression). It supposedly has the capability to hide files and text, but it crashed every time we tried to hide information (files and text).

Steganography (Android) did not work with any of the 8 test carrier files that we tested. In the process of hiding information in a JPEG image, the program actually converts the cover image to a rather large PNG image. Since Facebook was then doing an upload to download file size ratio compression in the range (7 to 43) and converting a PNG image to a JPEG image, it was obvious that additional tests would not succeed.

In all cases where we were able to download the loaded JPEG carrier images from Facebook (even if we could not recover the hidden information), we observed that the downloaded carriers were not garbled so that steganography could be visually detected. Also, we were correct in our prediction that few steganography programs would produce loaded carrier images that could survive Facebook compression.

## V. JP HIDE & SEEK CAPACITY TESTING

In phase 4 of the experimentation, we achieved some amount of success only with JP Hide & Seek (JPHS), Steg, and Steghide. Since we achieved a 50% success rate with JPHS, it appeared to be the best available tool for conducting Facebook steganography.

In this fifth phase of experimentation, we tested the capacity of 100 JPEG carrier files to hide text and images using JPHS for steganography and Facebook for transmission. Fifty of the images were in the 960 resolution, and the other 50 were in the 2048 resolution. We used only carrier images that successfully passed the initial 1 byte payload test. The steps conducted in each experiment were:

- Create a text file of selected size or choose an image file of selected size
- Hide the text or image in the selected preprocessed carrier image using JPHS
- Test recovery using JPHS
- Upload the test carrier to a Facebook album with high quality setting
- Download the carrier image from Facebook
- Attempt to recover the hidden image or text

JPHS was able to recover all hidden texts and images prior to the carrier test files being uploaded to Facebook. Thus, all failures encountered can be attributed to Facebook compression. The success rates after using Facebook to transmit the carrier files decreased as the hidden payload increased. Fig. 1 shows the success rates for hiding text files of various sizes in the fifty 960 resolution test carriers. Fig. 2 shows the success rates for hiding text files of various sizes in the fifty 2048 resolution test carriers. Fig. 3 shows the success rates for hiding image files of various sizes in the fifty 2048 resolution test carriers. We counted only files and images recovered 100% intact as successes. There were a few recoveries we did not count because characters were changed in the recovered text or the recovered images were grainy or partially corrupted. Fig. 4 shows a sample corrupted hidden image along with its original copy. We had no successes for hiding images in the fifty 960 resolution test carriers.

So with the 960 resolution test carriers, we had a 90% text payload recovery success rate at 65 bytes of hidden text, and then the success rate dipped to 56% at 400 bytes of hidden text. Then the success rate climbed to 76% at 1024 bytes and gradually decreased with increased hidden text payloads. With the 2048 resolution test carriers, we had a high text recovery success rate up to 400 bytes.

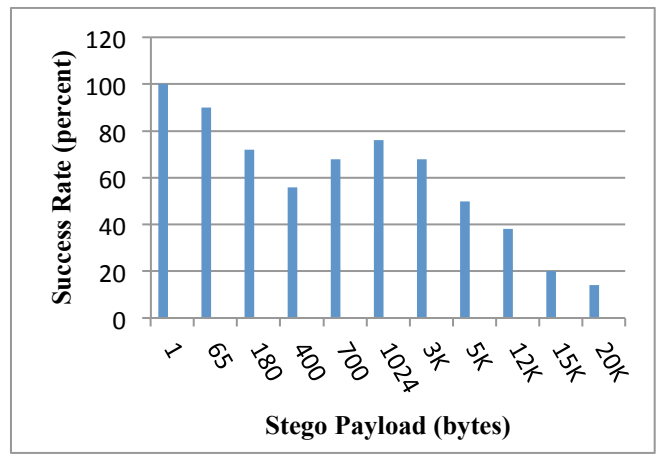

Figure 1. Success rates for hiding text in low resolution (960 pixel * yyy pixel) JPEG carriers.

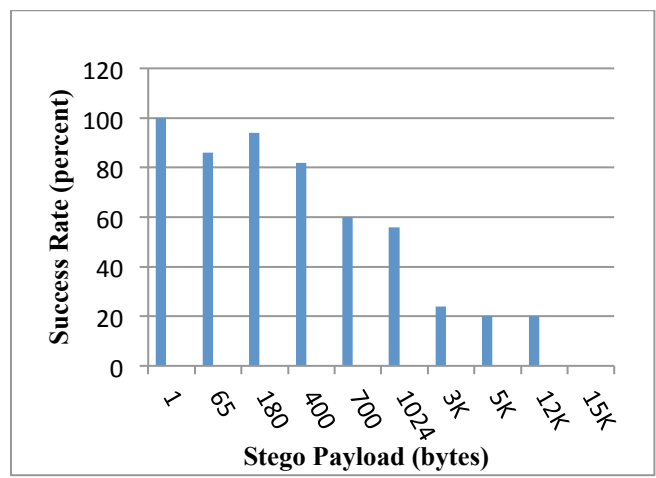

Figure 2. Success rates for hiding text in high resolution (2048 pixel * yyyy pixel) JPEG carriers.

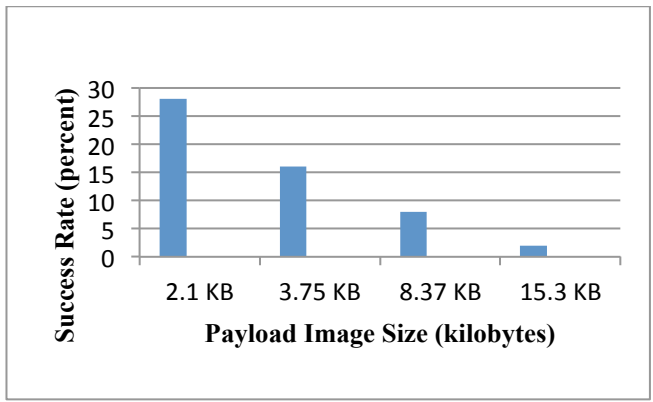

Figure 3. Success rates for hiding images in high resolution (2048 pixel * yyyy pixel) JPEG carriers.

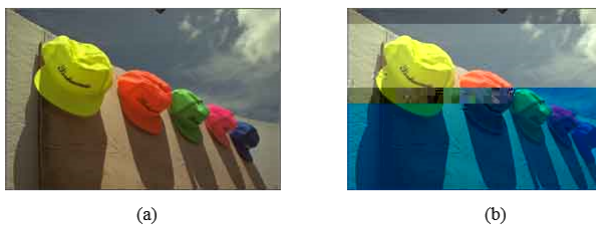

(a)                          (b)

Figure 4. (a) example of original image and (b) example of corresponding corrupted hidden image after hiding with JPHS, uploading to Facebook, downloading from Facebook, and extracting using JPHS.

Then the rate dropped off to around 60% at 700 and 1024 bytes and to about 20% at 3, 5, and 12 kilobytes. In both the high resolution and low resolution test cases, recovery failures and data corruption increased as we pushed more and more information relative to the carrier size through Facebook. Surprisingly, at many payload sizes the low resolution test carriers were better than the high resolution test carriers at successfully transmitting the hidden text. However, only the high resolution carriers were able to successfully transmit image payloads.

VI. STEGANALYSIS

*A. Overview*

Steganography can be detected by comparing the original file to the loaded cover file. However, in most real world scenarios the examiner will not have the original file. Steganography can also be detected with steganography program signatures. A steganography program's signature is found by observing repetitive changes to different original files when the program is used. Then, even when an examiner does not have the original file, the signature can be used to identify the program [13]. Most steganalysis programs use signature detection. Steganography can also be detected by examining the suspect file for statistical abnormalities. Means, variances, chi-square tests, linear analysis, Markov Fields, and wavelet statistics are examples of statistical examinations that can be done to measure the amount of departure from the expected norm and thereby detect distortion [12]. This is why steganalysis programs also typically employ statistical checks. In the specific case of JPEG steganography, Tech-faq.com [14] and Andriotis, Oikonomou, and Tryfonas [15] indicate that abnormal statistical distributions of JPEG coefficients can indicate hidden data.

*B. Applied to the Project*

We used a program called Steg Secret to check our loaded test carrier files before upload to Facebook and after upload to Facebook. Steg Secret comes in Spanish, so it may take a little work with a Spanish-English dictionary to get started. Of the steganography programs listed in section 4, Steg Secret was only able to detect steganography conducted with Our Secret. Steg Secret was able to detect Our Secret steganography before upload to Facebook, but it was not able to detect Our Secret steganography after download from Facebook. This is interesting because in this case the Facebook algorithm helps to further obscure the hidden information.

VII. CONCLUSIONS & FUTURE WORK

Facebook can be used to conduct JPEG steganography. Campbell-Moore developed the Secretbook Google Chrome plug-in to hide short text messages in JPEG carrier images and transmit them via Facebook. Although our model requires preprocessing work, multiple attempts, and testing to ensure success, we have shown that it is possible to hide longer text messages and small image files in JPEG cover files and transmit them using Facebook. Future work may include conducting more experiments to firm up the success rates, conducting more preprocessing to get better success rates, testing additional steganography tools, determining exactly what Facebook is doing to JPEG files during compression by detailed comparison of the uploaded and downloaded images, and developing a tool similar to Secretbook that can hide longer texts and files including images.

ACKNOWLEDGMENT

Krzysztof Szczypiorski was supported by the European Union in the framework of European Social Fund through the Warsaw University of Technology Development Programme.

REFERENCES

[1] Steganography. Retrieved February 11, 2015, from Wikipedia.org. http://en.wikipedia.org/wiki/Steganography.

[2] JPEG. Retrieved February 11, 2015, from Wikipedia.org. http://en.wikipedia.org/wiki/JPEG.

[3] A. Castiglione, G. Cattaneo, and A. De Santis, "A forensic analysis of images on online social networks," Proc. Third International Conference on Intelligent Networking and Collaborative Systems, 2011, pp. 679-684.
Retrieved February 11, 2015, from IEEE Xplore. http://ieeexplore.ieee.org.

[4] S. Nagaraja, A. Houmansad, P. Piyawongwisal, V. Singh, P. Agarwal, and N. Borisov, "Stegobot: A covert social network botnet," Proc. 13th International Conference on Information Hiding, 2011, pp. 299-313.
Retrieved February 12, 2015, from University of Massachusetts. https://people.cs.umass.edu/~amir/papers/IH11-Stegobot.pdf.

[5] R. Beckhusen, "Secretbook lets you encode hidden messages in your Facebook pics," Wired, April 10, 2013. Retrieved February 11, 2015, from wired.com. http://wired.com/2013/04/secretbook.

[6] BMP File Format. Retrieved February 11, 2015, from Wikipedia.org. http://en.wikipedia.org/wiki/BMP_file_format.

[7] Graphics Interchange Format. Retrieved February 11, 2015, from Wikipedia.org. http://en.wikipedia.org/wiki/GIF.

[8] Portable Network Graphics. Retrieved February 11, 2015, from Wikipedia.org. http://en.wikipedia.org/wiki/Portable_Network_Graphics.

[9] Tagged Image File Format. Retrieved February 11, 2015, from Wikipedia.org. http://en.wikipedia.org/wiki/Tagged_Image_File_Format.

[10] JPEG File Interchange Format. Retrieved February 11, 2015, from Wikipedia.org. http://en.wikipedia.org/wiki/JFIF.

[11] Exchangeable Image File Format. Retrieved February 11, 2015, from Wikipedia.org. http://en.wikipedia.org/wiki/Exif.

[12] G. Kessler, An overview of steganography for the computer forensics examiner. Retrieved February 11, 2015, from garykessler.net. http://www.garykessler.net/library/fsc_stego.html.

[13] P. Richer, "Steganalysis: Detecting hidden information with computer forensic analysis." Retrieved February 11, 2015, from SANS.org. http://www.sans.org/reading-room/whitepapers/stenganography/steganalysis-detecting-hidden-information-computer-forensic-analysis-1014.

[14] Steganography. Tech-faq.com, May 8, 2014. Retrieved December 4, 2014, from tech-faq.com. http://www.tech-faq.com/steganography.html.

[15] P. Andriotis, G. Oikonomou, and T. Tryfonas, "JPEG steganography detection with Benford's law," Digital Investigation, vol. 9, 2013, pp. 246-257. Retrieved February 11, 2015, from fortoo.eu. http://fortoo.eu/m/page-media/4/jpeg-steganography.pdf.